\documentclass[11pt]{article}
\usepackage[dvips]{graphicx}
\begin{document}
\tolerance=10000

%%\null

\hyphenpenalty=2000
\hyphenation{visco-elastic visco-elasticity}
\setcounter{page}{1}
\thispagestyle{empty}

%%\hoffset= 0.25truecm  %% TO RIGHT ->

%%%% FONTS

% ------------- specific fonts

\font\note=cmr10 at 10 truept  %% FOR ABSTRACT
\newcommand{\eproof}{\rule{0.2cm}{0.2cm}}

\newcommand{\stt}{\small\tt}
%%%%%%%%%%%  TITLE AND ABSTRACT %%%%%%

\font\title=cmbx12 scaled\magstep1

\font\bfs=cmbx10 scaled\magstep1
%%%%%%%%%%%%%%%%%%%%%%%%%%
%%%%  DEFINITIONS by F. MAINARDI  %%%%%%%%%%%%%%%%%%%%%%%%%%%%%%
\newcommand{\NN}{\bf{N}}
\newcommand{\ZZ}{\bf{Z}}
\newcommand{\CC}{\bf{C}}
\newcommand{\RR}{\bf{R}}
 \newcommand{\intl}{\int\limits}
\newcommand{\suml}{\sum\limits}

\def\pni{\par \noindent}
\def\vsh{\smallskip}
\def\vs{\medskip}
\def\vvs{\bigskip}
\def\vvvs{\bigskip\medskip} %% {\vskip 1.5truecm}
\def\vsp{\par}
\def\vsn{\vsh\pni}
\def\cen{\centerline}
\def\ra{\item{a)\ }} \def\rb{\item{b)\ }}   \def\rc{\item{c)\ }}
%%%%%%%%%%%%%%%%%%%
%%% DEF of BINOMIAL COEFFICIENTS
\def\alphak{{\alpha \choose k}}
\def\alphazero{{\alpha \choose 0}}
\def\alphaone{{\alpha \choose 1}}
\def\alphatwo{{\alpha \choose 2}}
\def\alphakk{{\alpha \choose k+1}}

\def\eg{{\rm e.g.}} \def\ie{{\rm i.e.}}
\def\sg{\hbox{sign}\,}
\def\sgn{\hbox{sign}\,}
\def\sign{\hbox{sign}\,}
\def\e{\hbox{e}}
\def\exp{\hbox{exp}}
\def\ds{\displaystyle}
\def\dis{\displaystyle}
\def\q{\quad}    \def\qq{\qquad}
\def\lan{\langle}\def\ran{\rangle}
\def\l{\left} \def\r{\right}
\def\lra{\Longleftrightarrow}
\def\arg{\hbox{\rm arg}}
\def\d{\partial}
 \def\dr{\partial r}  \def\dt{\partial t}
\def\dx{\partial x}   \def\dy{\partial y}  \def\dz{\partial z}
\def\rec#1{{1\over{#1}}}
\def\log{\hbox{\rm log}\,}
\def\erf{\hbox{\rm erf}\,}     \def\erfc{\hbox{\rm erfc}\,}
\def\zr{z^{-1}}    %%%% from Jerusalem

%%%% DEFINITION for the GREEN FUNCTION G_{\alpha,\beta}^\theta}
%%%% and their Fourier/Laplace transforms
\def\G{{G_{\alpha,\beta}^\theta}}
\def\Gzero{{G_{\alpha,\beta}^0}}
\def\K{K_{\alpha,\beta}^\theta}
\def\Kzero{K_{\alpha,\beta}^0}

\def\Gxt{\G (x,t)}
\def\Gkt{{\widehat{\G}}  (\kappa,t)}
\def\Gxs{{\widetilde{\G}}  (x,s)}
\def\Gks{{\widehat{\widetilde {\G}}} (\kappa,s)}
%%%%%%%%%%%%%%%%%%%%%%%%%%%%%%%%%%%%%%%%%%%%%%%%%%%%%%
\def\FT{{\cal F}\,} %% FOURIER TRANSF
\def\LT{{\cal L}\,}  %% LAPLACE TRANSF
\def\L{{\cal L}} %%% Laplace Transform !!!!
\def\F{{\cal F}} %%% Fourier Transform !!!!
\def\M{{\cal M}}  %%% Mellin Transform
%%\def\I{{\cal I}}  %% The generic open interval  \in \RR
%%%%%%%%\vfill\eject%%%%%%%%%%%

 \cen{FRACALMO PRE-PRINT   {\bf www.fracalmo.org}}

\vs

\hrule

\vsh
\begin{center}
%%  {\bf First International Conference on
%%  \\  "Nonlinear Analysis and Mechanics of Continuous Media"
%% \\  Ho Chi Minh City, Vietnam, 7-10 January 2003.}\\

{\bfs Fractional Diffusion Processes:}
\vs

{\bf Probability Distributions and Continuous Time Random Walk}\footnote{%%
Invited chapter to the book
{\it Processes with Long Range Correlations}, edited by G. Rangarajan and M. Ding, 
Springer-Verlag, Berlin 2003, pp. 148-166.
    [Lecture Notes in Physics, No. 621].
 The present e-print is a revised
 version (with up-date annotations and references) of that  contribution,
 but essentially represents our knowledge of that early time.}

\vvs

{Rudolf GORENFLO}$^{(1)}$ and 
 {Francesco MAINARDI}$^{(2)}$

\vs
%%%%%%%%%%%%%%%%
$\null^{(1)}$ Department of Mathematics and Computer Science,\\
	Freie Universit\"at  Berlin,
Arnimallee  3, D-14195 Berlin, Germany. \\
{\tt gorenflo@mi.fu-berlin.de}
\\ [0.25 truecm]

$\null^{(2)}$ Department of Physics, University of Bologna, and INFN,\\
Via Irnerio 46, I-40126 Bologna, Italy. \\
{\tt francesco.mainardi@unibo.it} \ {\tt mainardi@bo.infn.it}
\end{center}
% If there are more authors at one institute, you should first
% use \author{...} for each author followed by \institute{...}.
\cen{\bf Abstract} %%% Abstract follows

\vskip 0.1truecm
\noindent
A physical-mathematical %% dynamics
approach to anomalous diffusion
may be  based on generalized diffusion equations
(containing derivatives of fractional order in space or/and time)
and related random walk models.
 By   the space-time fractional diffusion
equation we mean an evolution equation obtained from the standard linear
diffusion equation by replacing the second-order space derivative
with a Riesz-Feller derivative of order $\alpha \in (0,2]$
and skewness $\theta$ ($|\theta|\le\hbox{min}\,\{\alpha ,2-\alpha \}$),
and the first-order time derivative with a Caputo derivative
of order $\beta \in (0,1]\,.$
The fundamental solution (for the {Cauchy} problem)
of the fractional diffusion equation  can be interpreted as
a probability density  evolving in time
of a peculiar {self-similar}  stochastic process.
We view it as a generalized diffusion process
that we call {\it fractional diffusion process},
and present an integral representation of the fundamental solution.
A more general approach to anomalous diffusion
is however known to be provided
by the master equation for a continuous time random walk (CTRW).
We show how this equation reduces to our
fractional diffusion equation
by a properly scaled passage to the limit of
compressed waiting  times  and  jump widths.
Finally, we describe a method of simulation and display
(via graphics) results of a few numerical case studies.

\vskip 2pt

\noindent
 {\it Mathematics Subject Classification 2000}:
%% {\it MSC 2000}:
26A33,  %%%%  (main);    Fractional derivatives and integrals
33E12, %% Mittag-Leffler type functions
33C60,  %% hypergeometric integrals and functions defined by them
44A10,  %% Laplace Transforms
45K05,  %% integro-partial differential equations
47G30,  %% Pseudo-differential operators
60G18, %%%  60 is Stochastic processes; 60G18 = Self-similar processes
60G50, %% sums of independent random variables, random walks
60G51, %% processes with independent increments (Levy processes!)
60G55, %% stable processes
60J60, %% diffusion processes
60J70. %% applications of diffusion processes

% \vskip 2pt

% \noindent
% {\it Key Words and Phrases}:
% Continuous time random walks,  anomalous diffusion, fractional derivatives,
% Mellin-Barnes integrals, stable  distributions, self-similar stochastic processes.

\newpage

%%%%%%%% SECTION 1 INTRODUCTION
\section{Introduction}
%% Start your manuscript with an introduction....

It is well known that the fundamental solution (or {\it Green function})
for the {Cauchy} problem of the   linear diffusion equation
can be interpreted as a Gaussian (normal)  probability density function ($pdf$)
in space, evolving in time.
All the moments of this $pdf$ are finite;
in particular, its variance
is proportional to the first power of time, a noteworthy property
of the {\it standard diffusion} that can
be understood by  means of an unbiased random walk model for
the {\it Brownian motion}.
%%%%%%%%

%%%%%%%%%%%

In recent years a number of master equations have been proposed
for random walk models that turn out to be
beyond the classical Brownian motion,
see \eg Klafter {\it et al}. \cite{Klafter PhysToday96}. %%   (1996).
In particular, evolution equations containing fractional derivatives
have gained revived interest in that they are expected to provide
suitable mathematical models  for describing phenomena of
anomalous diffusion, strange kinetics%%
\footnote{To the topic of strange kinetics  a special issue
(nowadays in press)
of {\it Chemical Physics} is  devoted where the
interested reader can find a number of
applications of fractional diffusion equations}
%%%%%%%%%%%%%%%%%%%%%%%%%%%%
and transport dynamics
in complex systems.
Recent references include \eg
\cite{%
%% Barkai PRE01,%%
Barkai CHEMPHYS02,%
BarkaiMetzlerKlafter PRE00,%%
ChechkinGonchar JETP00,%
GorMai NLD02,%
GorMai CHEMPHYS02,%
Hilfer 00a,%%
% HilferAnton 95,%
%% Mainardi WASCOM93,Mainardi CHAOS96,
Mainardi CISM97,%
Mainardi LUMAPA01,%%
MetzlerKlafter PhysRep00,%%
%% Metzler PhysA94,
Metzler CHEMPHYS02,%
Paradisi PhysA01,%
SaichevZaslavsky 97,%
%% SchneiderWyss 89,%
%% Uchaikin 00,%
UchaikinZolotarev 99}.%%
%% WestGrigoliniMetzlerTheo PRE97}.
%%% and references therein.

The paper is divided as follows.
%%%%%%%%%%%
In Section 2 %%% In this paper we first
we introduce
our    fractional diffusion equations
providing the reader with the essential notions for
the derivatives of fractional order (in space and in time)   entering
these equations.
More precisely, we  replace in the standard linear diffusion equation
the second-order space derivative or/and the first-order time derivative
by    suitable {\it integro-differential} operators, which can be
interpreted as a space or time derivative of fractional order
$\alpha \in (0,2]$ %% ($ 0< \alpha < 2$)
or $\beta \in (0,1]\,,$ %% ($0<\beta < 1$),
respectively%%%%
\footnote{We remind that the term "fractional" is a misnomer
since the order can be a real number and thus is not restricted to  be
rational. The term is kept only for historical reasons,
see \eg \cite{GorMai CISM97}.
Our fractional derivatives are required to
coincide with the standard derivatives of integer order as soon as
$\alpha=2$ (not as $\alpha =1$!)   and $\beta =1\,. $}.
%%%%%%% THE END OF THE FOOTNOTE %%%%
The space fractional derivative
is required to depend also on a real parameter $\theta$
(the {\it skewness})  subjected to the
restriction $|\theta|\le\hbox{min}\,\{\alpha ,2-\alpha \}\,. $
%%%%%%%%%%
Then, in Section 3 we pay attention to the fact that
the fundamental solutions (or {\it Green functions}) of our
diffusion equations      %% for the {Cauchy} problem)
of fractional order in space or/and in time
can be interpreted as   spatial  probability densities evolving in
time,   related to  certain {\it self-similar} stochastic process.
We view these processes as generalized (or {\it fractional})
diffusion processes to be properly understood through
suitable random walk models
that have been treated by us  in previous papers, see \eg
\cite{GorDFMai PhysA99,GorMai FCAA98,GorMai ZAA99,GorMai CHEMNITZ01,%
GorMai NLD02,GorMai CHEMPHYS02}.
%%%%%%%%
In Section 4  we show how
such evolution equations of fractional order
%% both in space and in time
can be  obtained from a more general  master equation
which governs  the so-called  continuous time
random walk (CTRW) by a properly scaled passage to the limit of
compressed waiting  times  and  jump widths.
%%%%%%%%%
The CTRW structure immediately offers a method of simulation
that we roughly describe in
Section 5 where we also display
 graphs  of a few numerical case studies.
%%%%%%%%%
Finally, in Section 6, we point out the main conclusions
and outline the direction for future work.

\section{The  space-time fractional diffusion  equation}

By  replacing in the standard diffusion equation
$$ {\d\over \dt} u(x,t) =  {\d^2\over \dx^2}\,
  u(x,t)\,,
   \q -\infty< x <+\infty\,, \q t \ge 0\,,
   \eqno(2.1)$$
where $u=u(x,t)$ is the (real) field variable,
%%%%%
the second-order space derivative and the first-order time derivative
by    suitable {\it integro-differential} operators, which can be
interpreted as a space and time derivative of fractional order
%% $\alpha$ ($ 0< \alpha \le 2$) or $\beta$ ($0<\beta \le 1$),%% respectively,
we  obtain a   generalized diffusion equation which may be
referred
to as the {\it space-time-fractional} diffusion equation.
We write this equation as
$$
_tD_*^\beta \, u(x,t) \,       = \, _xD_\theta^\alpha \,u(x,t) \,,
\q -\infty< x <+\infty\,, \q t \ge 0\,,
\eqno(2.2) $$
where  the $\alpha \,,\,\theta\,,\, \beta $ are real parameters
restricted as follows
$$ 0<\alpha\le 2\,,\q |\theta| \le \hbox{min} \{\alpha, 2-\alpha\}\,,
  \quad 0<\beta \le 1\,.\eqno(2.3)$$
In (2.2)
$\,_xD_\theta^\alpha \,$ is
the space-fractional
{\it Riesz-Feller derivative}  of order $\alpha $ and skewness $\theta\,,$
and  $\,_tD_*^\beta\,$  is
 the time-fractional {\it Caputo derivative} of order $\beta \,.$
The definitions of these fractional derivatives
are more easily understood if given
in terms of Fourier transform and Laplace transform, respectively.
%%%%%%%%%  Suggested by RG
Generically, $u(x,t)$ is interpreted as a mass density
or a probability density depending on the space variable $x$,
evolving in time $t$.

In terms of the Fourier transform we have  for the
space-fractional {\it Riesz-Feller derivative}
$$ {\cal F} \l\{\, _xD_\theta^\alpha\, f(x);\kappa \r\} =
  - \psi_\alpha ^\theta(\kappa ) \,
  \, \widehat f(\kappa) \,,
\q
   \psi_\alpha ^\theta(\kappa ) =
|\kappa|^\alpha \, \e^{\ds  i (\sgn \kappa)\theta\pi/2}\,,
%% \q 0<\alpha  \le 2\,, \q
%% |\theta| \le  \,\hbox{min}\, \{\alpha ,2-\alpha \}\,.
%% \q  \theta \in \re\,.
\eqno(2.4)$$
 where
$  \widehat f(\kappa)  =
{\cal F} \l\{ f(x);\kappa \r\}
  = \int_{-\infty}^{+\infty} \e^{\,\ds +i\kappa x}\,f(x)\, dx\,.$
%%%  \; \kappa \in \re\,.$
In other words the symbol of the pseudo-differential operator
$\,_xD_\theta^\alpha$ is required to be the logarithm of the
characteristic function of the generic  {\it stable}
(in the L\'evy sense)
probability density, according to the Feller parameterization
\cite{Feller 52,Feller 71}.
 We note that the allowed region for the %%% real
parameters $\alpha $ and $\theta$
turns out to be
 a {diamond} in the plane $\{\alpha \,,\, \theta\}$
with vertices in the points
$(0,0)\,, \,(1,1)\,,\,(2,0)\,,\, (1,-1) \,,$
that we call the {\it Feller-Takayasu diamond}%%
$\,$\footnote{Our notation  for the
stable distributions has been adapted
from the original one by Feller. From 1998,
see \cite{GorMai FCAA98}, we have found it as the
most convenient among the others available in the literature, see \eg
\cite{JanickiWeron 94,Levy STABLE,MontrollShlesingher 84,MontrollWest 79,%%
SamoTaqqu 94,Sato 99,UchaikinZolotarev 99}.
%% Janicki \& Weron \cite{JanickiWeron 94},
%% L\'evy \cite{Levy STABLE},
%% Montroll and associates \cite{MontrollShlesingher 84,MontrollWest 79},
%% Samorodnitsky \& Taqqu \cite {SamoTaqqu 94},
%% Sato \cite{Sato 99},
%% Uchaikin \& Zolotarev \cite{UchaikinZolotarev 99}.
%%% Zolotarev \cite{Zolotarev 86}.
Furthermore, this notation
has the advantage that  the whole class of the {\it strictly stable}
densities is represented. As far as we know,
the diamond representation in the plane
$\{\alpha ,\theta\}$ was formerly given by Takayasu
in his 1990 book on {\it Fractals}
\cite{Takayasu FRACTALS}.}. %% see FCAAfig1.eps in LUMAPA paper !
%%%%%%%%
\noindent
For $\alpha =2$ (hence $\theta=0$) we have
$ {\cal F} \l\{\, _xD_\theta^\alpha\, f(x);\kappa \r\} =  -\kappa ^2
= (-i\kappa )^2\,,$ so
we recover the standard
second derivative. More generally for $\theta=0$
we have
$ {\cal F} \l\{\, _xD_\theta^\alpha\, f(x);\kappa \r\} =
              -|\kappa |^\alpha  = - (\kappa ^2)^{\alpha /2}$ so
$$ \,_xD_0^\alpha  = - \l(-{d^2\over dx^2}\r) ^{\alpha/2}\,.
     \eqno(2.5) $$
In this case we refer to the LHS of (2.5) as simply to the
{\it Riesz fractional derivative} of order $\alpha \,.$
%%%%%%%%%%%%%%%%%%%%
Assuming $\alpha \ne 1,2$ and taking $\theta$ in its range
one can show that the explicit expression of the
{\it Riesz-Feller fractional derivative} obtained from (2.4)
is
  $$ \,_x D_\theta ^\alpha \, f (x) :=
  -\, \l[
  c_+(\alpha,\theta)\,\,_x  D^{\alpha}_+
+ c_-(\alpha,\theta)\,\,_x  D^{\alpha}_-
     \r]\,  f(x)  \,,  \eqno(2.6)$$
where, see \cite{GorMai FCAA98},
$$
c_+(\alpha,\theta) =
  {\sin \,\l[(\alpha-\theta)\,\pi/2\r] \over\sin\,(\alpha\pi)}\,,
 \qq
   c_-(\alpha,\theta) =
{\sin\,\l[(\alpha+\theta)\,\pi/2\r] \over \sin(\alpha\pi)} \,,
\eqno(2.7)
$$
%%%%%%%%%%%
% \vfill\eject
%%%%%%%%%%%
\noindent
and $_xD_\pm^{\alpha} $ are Weyl fractional derivatives
defined as
 $$
 _xD_\pm^{\alpha} \,f(x) = \cases{
     {\ds \pm {d \over dx}} \,\l[\,_x I_\pm^{1-\alpha}\,f(x)\r] \,,
   & if $\q 0<\alpha < 1 \,,$\cr\cr
     {\ds{d^2 \over dx^2}} \,\l[\, _x I_\pm^{2-\alpha}\,f(x)\r]
 \,,
   & if $\q 1<\alpha < 2 \,.$\cr}
\eqno(2.8)
$$
In (2.8) the $\,_x I_\pm^\mu $ ($\mu >0$) denote the Weyl fractional
integrals defined as
 $$ \cases{{\ds _x I_+^\mu  \, f(x)}=
 {\ds \rec{\Gamma(\mu )}}\,
{\ds   \int_{-\infty}^x \!\!  (x-\xi)^{\mu -1}\, f(\xi)\,d\xi} \,,
 \cr\cr
 {\ds _x I_-^\mu  \, f(x)}=
 {\ds \rec{\Gamma(\mu )}}\,
 {\ds  \int_x^{+\infty} \!\! (\xi-x)^{\mu -1}\,f(\xi)\,d\xi} \,.
  \cr} \; (\mu >0) \eqno(2.9)$$
In the particular case $\theta =0$ we get
$c_+(\alpha ,0)  = c_-(\alpha ,0) =
  {1 /[2 \cos\,(\alpha\pi /2)}] \,,$
and, by passing to the limit for $\alpha  \to 2^-\,,$
we get $\, c_+(2,0) = c_-(2,0) = - 1/2\,. $

For $\alpha =1$ we  have
$$ _xD^1_\theta \, f(x) = \l[ \cos (\theta \pi/2)\,  _xD_0^1
 + \sin (\theta \pi/2)\,  _xD \r] \, f(x) \,,\eqno(2.10)$$
where
$_xD \, f(x) = {\ds {d\over dx}}\, f(x)\,,$ and
$$
  _xD_0^1 \,f(x) = -{d\over dx} \,\l[ _xH\, f(x)\r] \,,
  \q _xH \, f(x) =  {1\over \pi}
\, \l( \int_{-\infty}^{+\infty} {f(\xi )\over x-\xi}\, d\xi\r)
%% =    {1\over \pi}
%%\, \l( \int_{-\infty}^{+\infty} {f(x-\xi )\over \xi}\, d\xi\r)
 \,.
\eqno(2.11) $$
In (2.11) the operator $\,_xH$ denotes the Hilbert transform
and its singular integral is understood in the Cauchy principal value
sense, see \cite{GorMai CHEMNITZ01}.

The operator $_xD^\alpha_\theta$
%% defined through (2.6)-(2.11)
has been referred to as
the {\it Riesz-Feller} fractional derivative since
both Marcel Riesz and William Feller contributed to its definition%%%
\footnote{%%%%%%%%%  FOOTNOTE on RIESZ and FELLER
Originally,   in the late  1940's, Riesz \cite{Riesz 49}
introduced  the pseudo-differential operator $_x I_0^\alpha$ whose
 symbol is $|\kappa|^{-\alpha} \,,$  well defined for any
 positive $\alpha$ with the exclusion of odd integer numbers,
 afterwards named the {\it Riesz potential}.
  The Riesz fractional derivative $_x D_0^\alpha := - \,_x I_0^{-\alpha}$
 defined by analytical continuation was generalized by Feller
 in his 1952 genial paper \cite{Feller 52} to include the skewness
 parameter      of the strictly stable densities.}.

Let us now consider the time-fractional {\it Caputo derivative}.
Following the original idea by Caputo
 \cite{Caputo 67},  %% ,Caputo 69},
see also
\cite{ButzerWestphal 00,CaputoMaina 71,GorMai CISM97,Podlubny 99},
a proper time fractional derivative of order $\beta \in (0,1)\,,$
useful for physical applications, may be defined in terms
of the following  rule for its Laplace transform%
%%%%%%%%%%% FOOTNOTE ON REAL PARAMETER in Laplace Transf %%%
\footnote{For our purposes we agree to take the Laplace
parameter $s$ real}
%%%%%%%%%%%%%%%%%%%
$$ {\cal L} \l\{ _tD_*^\beta \,f(t) ;s\r\} =
      s^\beta \,  \widetilde f(s)
   -    s^{\beta  -1}\, f(0^+) \,,
  \q 0<\beta  < 1 \,, \eqno(2.12)$$
where
$ \widetilde f(s) =
{\cal L} \l\{ f(t);s\r\}
 = \int_0^{\infty} \e^{\ds \, -st}\, f(t)\, dt\,.$
%% \; s \in \CC\,. $
 Then
the {\it Caputo fractional derivative} of
%% a sufficiently well-behaved function
$f(t)$  turns out to be defined as
$$
    _tD_*^\beta \,f(t) :=
    {\ds \rec{\Gamma(1-\beta )}}\,{\ds\int_0^t
 {\ds {f^{(1)}(\tau)\, d\tau \over (t-\tau )^{\beta  }}}} \,,
 \q 0<\beta  <1\,.\eqno(2.13)$$
In other words the  operator
$\,_tD_*^\beta $ is required to generalize the well-known
rule for the Laplace transform of
the first derivative of a given (causal) function keeping the standard
initial value of the function itself%%
%%%%% FOOTNOTE on R-L and CAPUTO fractional derivatives
\footnote{
The reader should observe that the {\it Caputo} fractional derivative
differs from the usual {\it Riemann-Liouville} fractional derivative
which, defined as the left inverse of the Riemann-Liouville fractional
integral, is  here denoted as
$\, _tD^\beta \,f(t)\,. $
We have,  see \eg \cite{SKM 93},
$$
 _tD^\beta  \,f(t) :=
   {d \over dt}\,\l[
   \rec{\Gamma(1-\beta)}\,\int_0^t
    {f(\tau)\,d\tau  \over (t-\tau )^\beta}\r] \,,\q 0<\beta <1\,. $$
It turns out that
$$ _tD_*^\beta  \,f(t)  \, = \, _tD^\beta  \,\l[ f(t) -
   f(0^+) \r] =
\, _tD^\beta  \, f(t) -
      f(0^+) \,
{t^{-\beta }\over \Gamma(1-\beta)}\,,
\q       0 <\beta <1\,.  $$
The {\it Caputo} fractional derivative,
 practically ignored in the  mathematical treatises,
represents a sort of regularization in the time origin for the
{\it Riemann-Liouville} fractional derivative
and  satisfies the  relevant property
of being zero when applied to a constant.
For more details on this fractional derivative (and its
extension to higher orders) we refer the interested reader to
Gorenflo and Mainardi \cite{GorMai CISM97}.}.
%%%%%% THE END of FOOT NOTE on CAPUTO %%%%5

%%  The generalized diffusion equation in (2.2) is referred to as
The {\it space-time fractional diffusion} equation (2.2)
 contains as particular cases
the {\it strictly space fractional diffusion} equation
    when $0<\alpha<2$ and $\beta =1\,,$
the {\it strictly time fractional diffusion} equation
   when $\alpha=2$ and $0<\beta<1\,,$
  and
the {\it standard diffusion equation} (2.1) when $\alpha=2$ and $\beta =1\,.$

For the equation (2.2) we consider the
Cauchy problem
$$
u(x,0^+)=   \varphi(x)\,,\q x\in {\RR} \,,\qq
 u(\pm\infty,t)=   0\,,\q t>0\,, \eqno(2.14) $$
where $\varphi(x)$ is a sufficiently well-behaved function.
By its solution
%% of the Cauchy   problem for the equation (2.2)
we mean a function $u_{\alpha, \beta}^\theta(x,t)$
 which satisfies the conditions (2.14).
By its  Green function (or fundamental solution)
we mean the (generalized)
function $G_{\alpha, \beta}^\theta(x,t)$
which, being the formal solution of (2.2) corresponding
to $\varphi(x) = \delta (x)\, $   (the Dirac delta function),
allows us to represent the solution of the Cauchy problem by
the integral formula
$$ u_{\alpha, \beta}^\theta(x,t) =
\int_{-\infty}^{+\infty} G_{\alpha, \beta}^\theta(\xi ,t)\,
 \varphi (x-\xi) \, d\xi \,. \eqno(2.15)$$
It is straightforward to derive  from (2.2) and (2.14) the
composite Fourier-Laplace transform of the Green function by taking
into account the Fourier transform for the {\it Riesz-Feller}
space-fractional derivative, see (2.4),
and the Laplace transform for the {\it Caputo}
time-fractional derivative,
see (2.12).
We have (in an obvious notation)
 $$- \psi_\alpha^\theta(\kappa ) \,\Gks  \,= \,
 s^\beta \,\Gks - s^{\beta -1}    \,,
\eqno(2.16)  $$
%% where
%% $$ \psi_\alpha^\theta (\kappa) :=
%%  |\kappa|^\alpha \, \e^{\ds  i (\sgn \kappa)\theta\pi/2}
%%   = \overline{ \psi_\alpha^\theta (-\kappa)}
%%  =  \psi_\alpha^{-\theta} (-\kappa)  \,. $$
so that
$$
  \Gks = {s^{\beta -1} \over
 s^\beta + \psi_\alpha^\theta (\kappa)} \,,
 \q s > 0\,,\q \kappa \in \RR\,.
 \eqno(2.17)$$
%% \noindent
In the special  case  $\theta =0$ we get
$$  \widehat{\widetilde{\Gzero}}(\kappa ,s)
    =  { s^{\beta -1} \over s^\beta + |\kappa |^\alpha }\,,
  \q s > 0\,,\q \kappa \in \RR\,.
    \eqno(2.18)$$
By using the known scaling rules
for the Fourier  and Laplace  transforms,
we infer directly from (2.17)
(without inverting the two transforms)   the  following
 noteworthy {\it scaling property}
of the Green function,
$$ \G(x,t)  =
    t^{-\beta /\alpha}\,\K \l(x/t^{\beta/\alpha}\r)\,.
   \eqno(2.19)     $$
Here $ x/t^{\beta/\alpha}\,$ acts as the {\it similarity variable}  and
$\K (\cdot)$ as  the {\it reduced Green function}.
%% We easily note that
Using (2.19) and the initial condition
$\G(x, 0^+) = \delta (x)\,,$ we note that
$$ \int_{-\infty}^{+\infty} \G(x,t) \, dx \,=\,
       \int_{-\infty}^{+\infty} \K(x) \, dx \, \equiv   1 \,.
%% \, \hbox{const}\,.
%%     \widehat{\K} (0) =1\,.
\eqno(2.20)$$
In the case of the standard diffusion equation (2.1)
the Green function is nothing but
the Gaussian probability density %% function
with variance
$\sigma^2  =2t\,,$ namely
$$G_{2,1}^0 (x,t)
 = {1\over 2\sqrt{\pi }}\,t^{-1/2}\, \e^{-\ds x^2/(4t)}\,.
\eqno(2.21)$$
In the general case,
following the arguments by Mainardi, Luchko \& Pagnini
\cite{Mainardi LUMAPA01},
 we can prove that $\G(x,t)$ is still a probability
density evolving in time.
%% For this it will be sufficient
%% to show that the reduced Green function is non-negative.
In the next section we summarise some results from
\cite{Mainardi LUMAPA01}.
%% Future suggestion to  justify for the fractional cases
%% the word "diffusion".  There are  TWO ASPECTS OF DIFFUSION
%% MACROSCOPIC aspect: redistribution of mass in proceeding time,
%%                   mass density evolving in time.
%% MICROSCOPIC aspect: sample paths of particles, sojourn probability density
%% The discrete-discrete model allows both interpretations

\section{The Green function for  space-time fractional diffusion}

%%%%%%%%%%
For the analytical and computational determination of
the  reduced Green function, from now on
we  restrict our attention  to $x>0\,$
because of the
{\it symmetry relation} %% that
$ K_{\alpha ,\beta}^\theta(-x) = K_{\alpha ,\beta}^{-\theta}(x)
\,.$
Mainardi, Luchko \& Pagnini \cite{Mainardi LUMAPA01} have provided
(for $x>0$) the Mellin-Barnes integral representation
$$ \qq \qq \K(x) =
{1\over  \alpha x}
{1\over 2\pi i} \int_{\gamma-i\infty}^{\gamma+i\infty}
{\Gamma({s\over \alpha}) \, \Gamma(1-{s\over \alpha}) \,\Gamma(1-s)
 \over \Gamma(1-{\beta\over \alpha}s) \,
 \Gamma ( \rho \,s)\,
 \Gamma (1-\rho \,s)}
 \, x^{\,\ds s}\,  ds
\,, \qq \qq  \eqno(3.3)  $$
$$ \rho =   { \alpha -\theta \over 2\,\alpha }\,,$$
where $0< \gamma < \hbox{min} \{\alpha ,1\}. $
%%\vfill\eject%%%%%%%%%%%
Following   \cite{Mainardi LUMAPA01},
we note that the
 Mellin-Barnes%%
%%% FOOTNOTE on  PINCHERLE  %%%
\footnote{
The names Mellin and Barnes refer to the two authors,
who in the first 1910's
developed the theory of these integrals  using them
for a complete integration of the hypergeometric differential equation.
We note that, as pointed out  in \cite{Erdelyi HTF}
(Vol. 1, Ch. 1, \S 1.19, p. 49), these integrals were first used
by S. Pincherle in 1888.
For a revisited analysis of the pioneering work of Pincherle
%% (Professor of Mathematics at the
%% University of Bologna from 1880 to 1928)
we refer to
%% the recent paper by Mainardi and Pagnini
\cite{MainardiPagnini OPSFA01}.}
%%%%%% THE END of the Footnote on PINCHERLE
 integral representation allows us to
 construct computationally   the fundamental solutions
of  (2.2) for any triplet $\{\alpha , \theta, \beta\}$
by  matching their
convergent and asymptotic expansions.
%%%%%
Readers acquainted with  Fox $H$ functions can recognize
in (3.3) the representation of a certain function of this
class, see \eg
\cite{Hilfer 00a,MathaiSaxena H,%%
%% Metzler PhysA94,
SKM 93,Schneider LNP86,%%
%% SchneiderWyss 89,%%
Srivastava H,UchaikinZolotarev 99}.
 Unfortunately, as far as we know, computing routines for this general
 class of special functions are not yet available.
%%%%%%%%%

Let us now point out the main characteristics of the peculiar cases
of {\it strictly space fractional diffusion} and
{\it strictly time fractional diffusion},
for which  the
 non-negativity of the corresponding reduced Green functions
 is known.
%%%%%%%%%%
For  $\beta =1$ and $0<\alpha <2$
({\it strictly space fractional diffusion})
%% we recover
 we have
 $$K_{\alpha, 1}^\theta (x) = L_\alpha^\theta(x) =
{1\over  \alpha x}
{1\over 2\pi i} \int_{\gamma-i\infty}^{\gamma+i\infty}
{\Gamma({s/ \alpha})  \,\Gamma(1-s)
 \over  \Gamma ( \rho \,s)\, \Gamma (1-\rho s)}
 \, x^{\,\ds s}\,  ds\,,\q x>0\,,\eqno(3.6) $$
with $ 0 <\gamma<\hbox{min}\{\alpha,1\}\,,$
where $L_\alpha ^\theta(x)$ denotes
the class of the
strictly stable (non-Gaussian)
densities exhibiting heavy  tails (with the algebraic decay
$\propto |x|^{-(\alpha +1)}$)
  and infinite variance.
%%%%%%
For $\alpha =2$ and $0<\beta <1$
({\it strictly time fractional diffusion})
%% we recover
 we have
$$K^\theta_{\alpha, 1} (x) = \rec{2} \, M_{\beta/2}(x)=
{1\over  2 x}
{1\over 2\pi i} \int_{\gamma-i\infty}^{\gamma+i\infty}
{\Gamma(1-s)
 \over   \Gamma (1- \beta s/2)}
 \, x^{\,\ds s}\,  ds\,,\q x>0\,,
\eqno(3.7) $$
with $0 <\gamma< 1\,,$
 where $\rec{2}\, M_{\beta/2}(x)$ denotes
the class
of the Wright-type densities exhibiting stretched exponential
tails   and therefore finite variance.
The corresponding Green function evolves
in time with the variance proportional to $t^\beta \,.$
Mathematical details
%% on these two classes       of probability densities
can be found
in \cite{Mainardi LUMAPA01};  for further reading
we refer   to Schneider \cite{Schneider LNP86} for stable densities,
and to Gorenflo, Luchko \& Mainardi \cite{GoLuMa 99}  %%% ,GoLuMa 00}
for the Wright-type densities.
%%%%%%%%%
For the special case $\alpha =\beta \le 1$
referred in \cite{Mainardi LUMAPA01} as
{\it neutral diffusion},
we obtain from (3.3) an elementary  (non-negative) expression
$$ \qq \qq \qq K_{\alpha,\alpha}^\theta (x) =
  N_\alpha ^\theta(x) =
{1\over  \alpha x}
{1\over 2\pi i} \int_{\gamma-i\infty}^{\gamma+i\infty}
{\Gamma({s\over \alpha}) \, \Gamma(1-{s\over \alpha})
 \over
 \Gamma ( \rho \,s)\,
 \Gamma (1-\rho \,s)}
 \, x^{\,\ds s}\,  ds \qq  \qq \qq
\eqno(3.8)$$
$$  =
{1\over  \alpha x}
{1\over 2\pi i} \int_{\gamma-i\infty}^{\gamma+i\infty}
 {\sin (\pi\, \rho \,s)
 \over
 \sin (\pi \,s/\alpha )}
 \, x^{\,\ds s}\,  ds
=     {{1\over\pi}}\,{x^{\alpha-1} \sin[{\pi\over 2}(\alpha -\theta )] \over
1 + 2x^\alpha \cos[{\pi\over 2}(\alpha -\theta)] + x^{2\alpha}}
\,, \q x>0\,,
     $$
with $0<\gamma <\alpha \,, $
 where $ N_{\alpha }^\theta(x)$ denotes
a peculiar class
of  densities exhibiting a power-law decay $\propto |x|^{-(\alpha+1) }\,,$
which contains the well known (stable) Cauchy  density
(recovered for $\alpha =1$ and $\theta =0$).
%%%%%%%%%

For the generic case of {\it strictly space-time diffusion}
($0<\alpha <2, \, 0<\beta <1$),
including neutral diffusion,  %%  for $\alpha = \beta <1$,
we can prove
%% Mainardi, Luchko \& Pagnini \cite{Mainardi LUMAPA01} have proven
the non negativity of the corresponding
reduced Green function
 in virtue of the   identity,  see  \cite{Mainardi LUMAPA01},
$$ \K(x) =
   \alpha   \, {\ds \int_0^\infty} \!\!
  \l[\xi ^{\alpha -1}\,  {M}_{\beta}\l(\xi ^{\alpha}\r)\r]\,
  {L}_{\alpha}^\theta\l({x/\xi }\r) \,
   {\ds{d\xi  \over \xi }} \,,\q 0<\beta <1\,, \q x>0\,.
 \eqno(3.9)$$
%%%%%%%%%%%%\vfill\eject%%%%%%%%%%%%%%
%% \noindent
 Then, as a consequence of the previous discussion,
 for the  {\it strictly space-time fractional} diffusion %% equation
 %% ($0<\alpha <2\,,\,0<\beta  < 1$)
 we obtain  a class of  probability densities
(symmetric or non-symmetric according to $\theta =0$ or $\theta \ne 0$)
which exhibit  heavy tails
with an algebraic decay $\propto |x|^{-(\alpha +1)}\,.$
Thus they belong to the domain of attraction of the L\'evy stable densities
of index $\alpha $ and  can be referred to as
{\it fractional stable densities}, according to
a terminology proposed by Uchaikin \cite{Uchaikin PC00}.
%%%%%%%%%%%%%
% \vfill\eject
%%%%%%%%%%%%
In Fig. 1 we exhibit some plots of the
probability densities provided by the
reduced Green function %% $\K(x)$
for some "characteristic" values of the parameters
$\alpha,\ \beta,$ and $\theta$.
These plots, taken from \cite{Mainardi LUMAPA01},
 were drawn %% by using the MATLAB system
for the values of the independent
variable $x$ in the range $|x|\le 5$.
To give the reader a better impression about
the behaviour of the tails the logarithmic scale was adopted.

\begin{figure}

\centerline{
 \includegraphics[width=.48\textwidth]{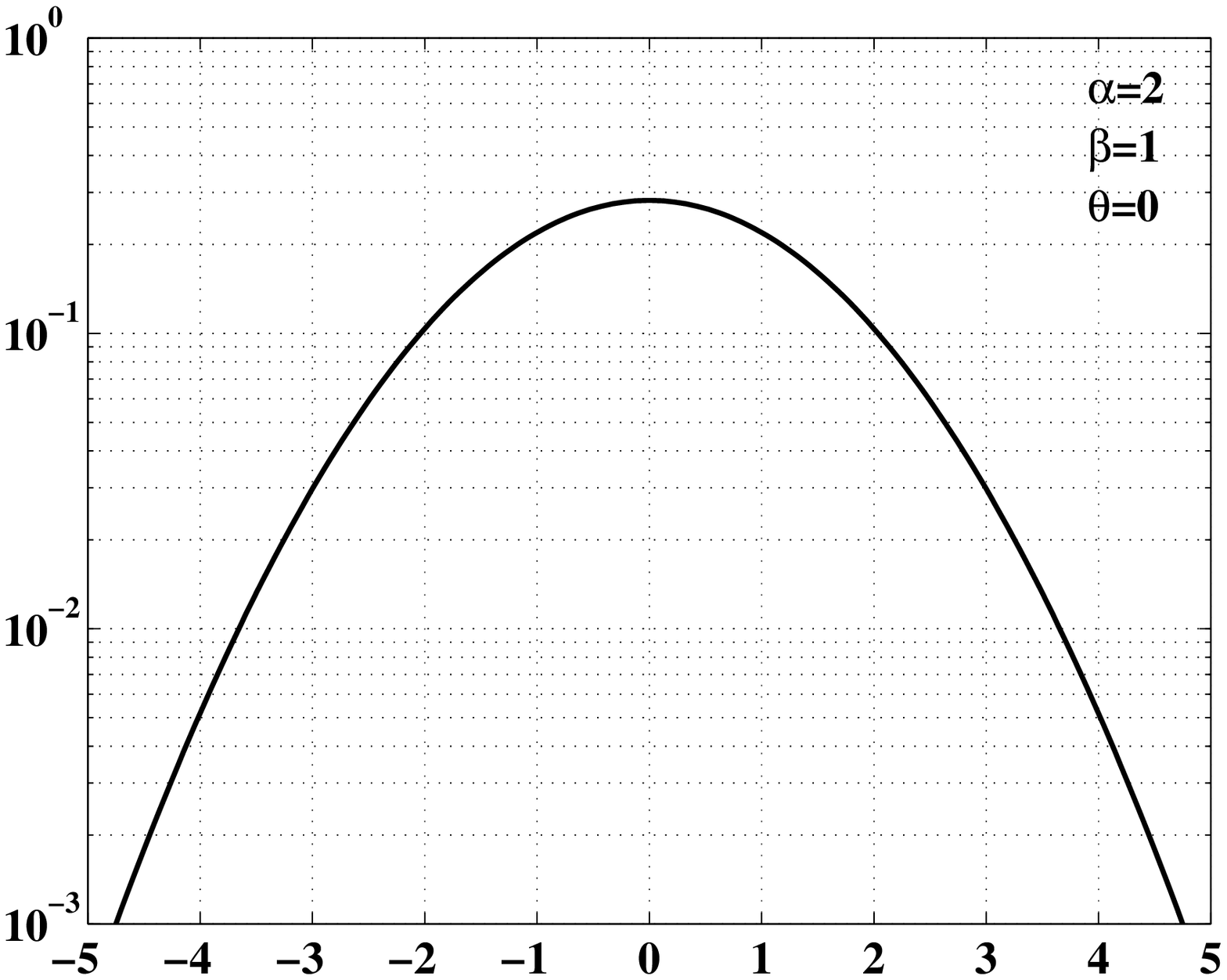}  %%FCAAfig2b.eps
 \includegraphics[width=.48\textwidth]{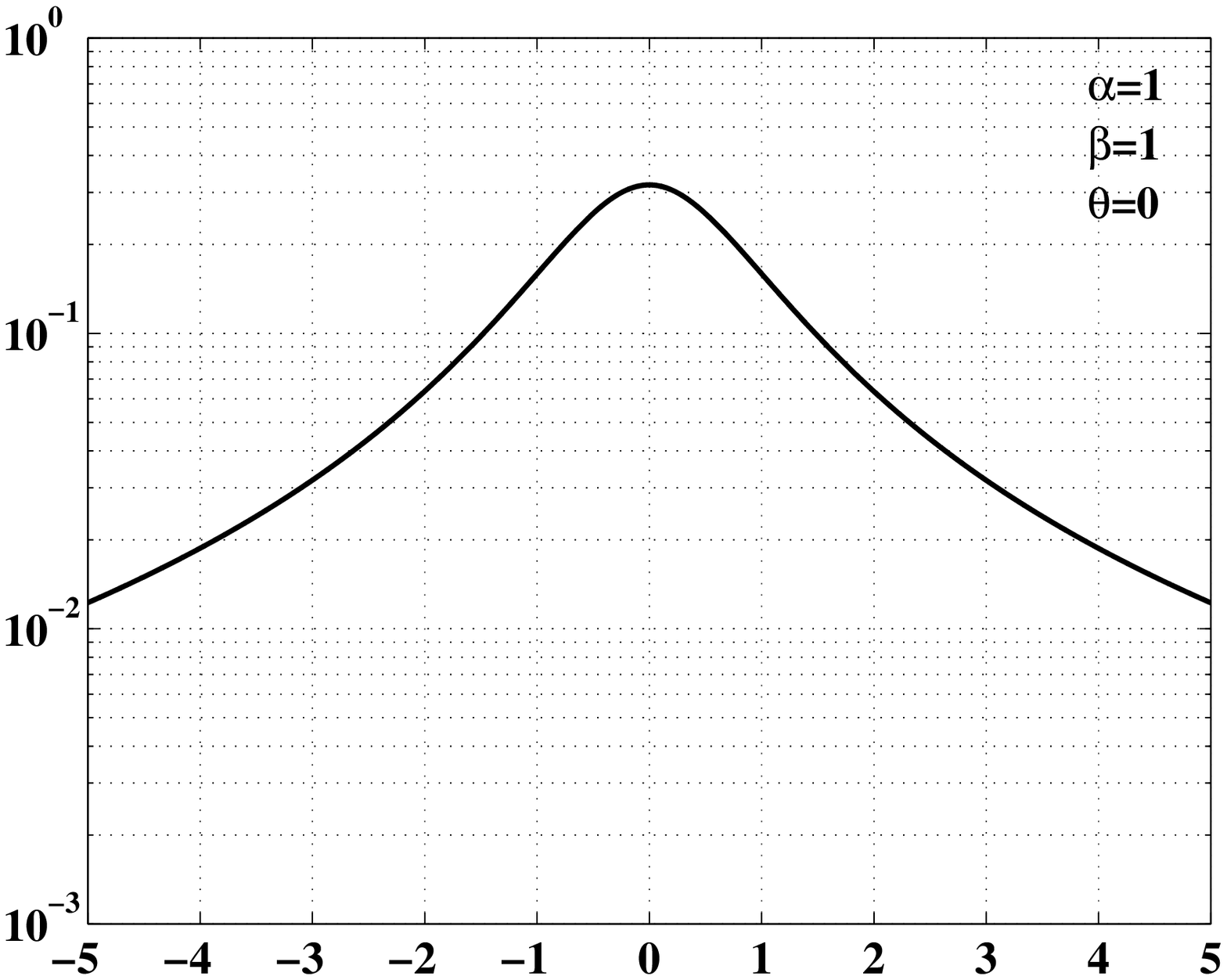}} %%FCAAfig4a.eps

\centerline{
 \includegraphics[width=.48\textwidth]{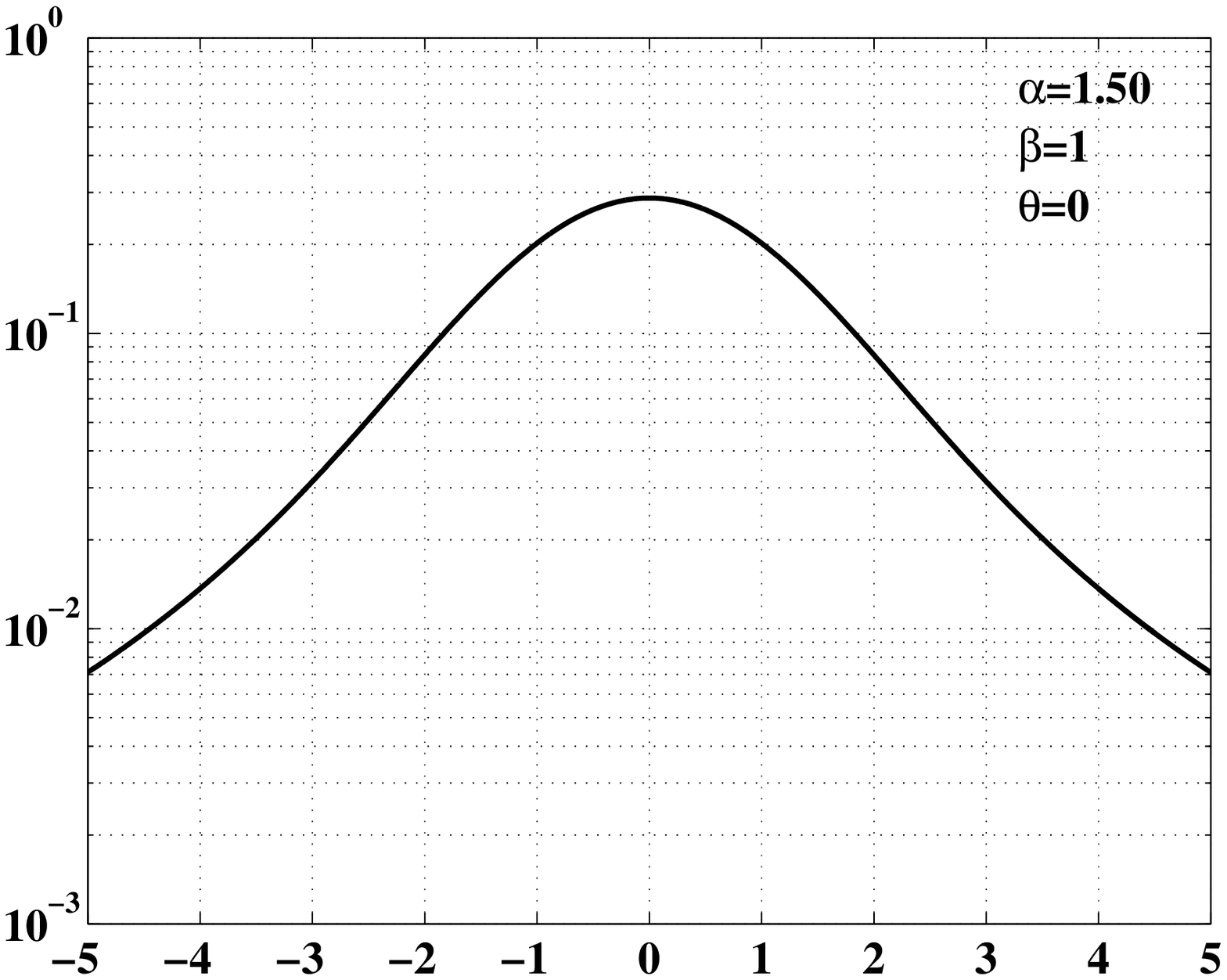}  %%FCAAfig5a.eps
 \includegraphics[width=.48\textwidth]{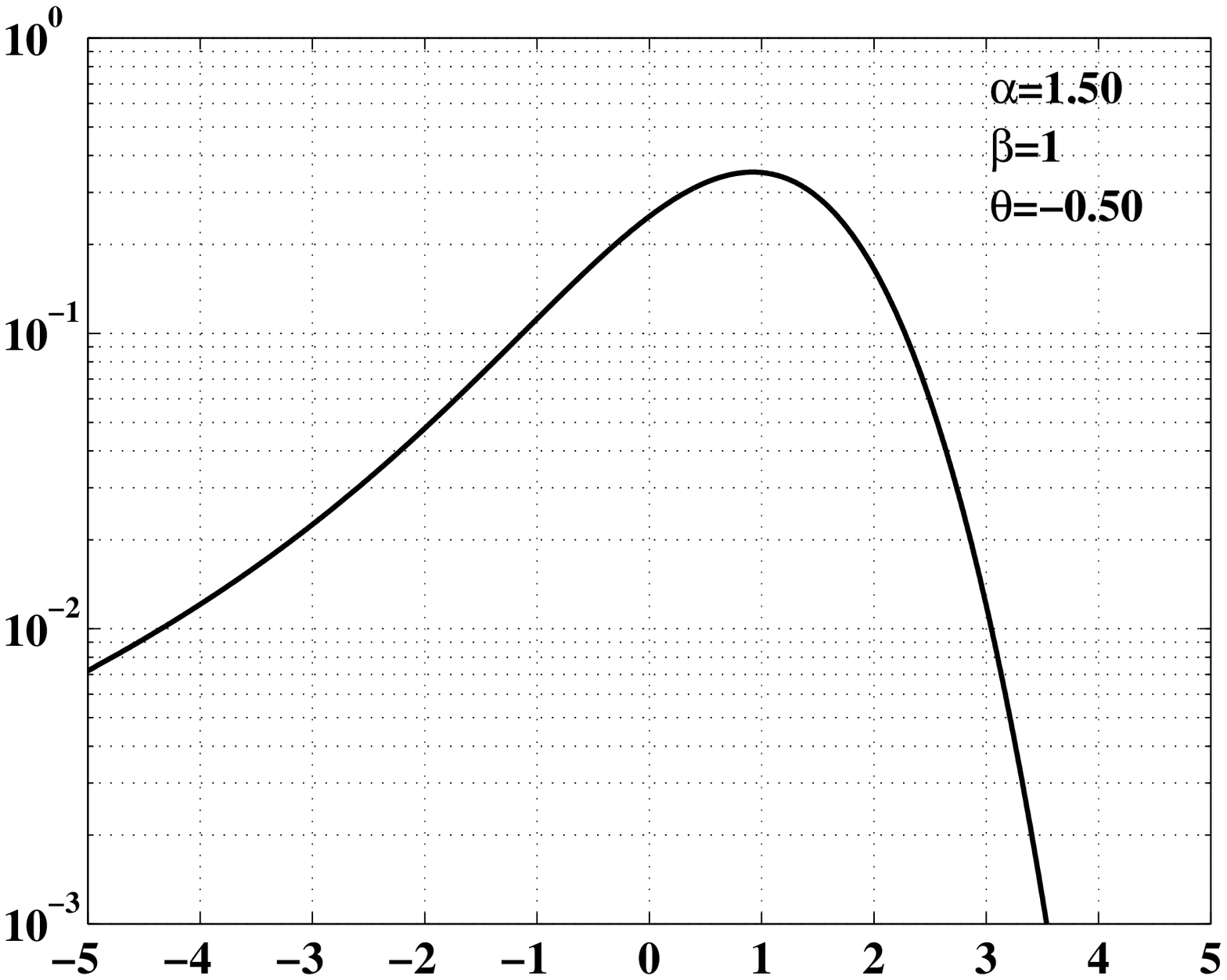}} %%FCAAfig5b.eps

\centerline{
 \includegraphics[width=.48\textwidth]{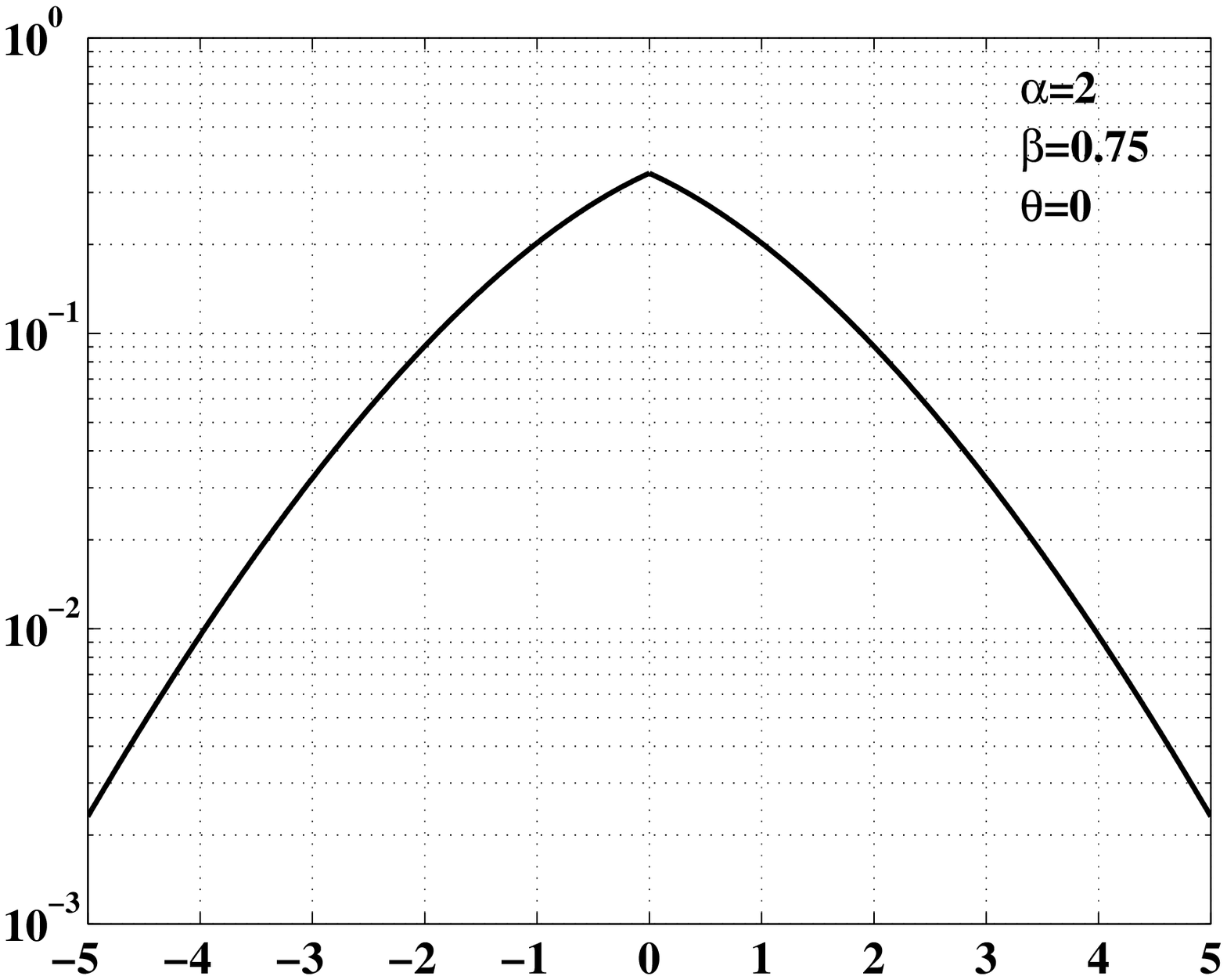}  %%FCAAfig7a.eps
 \includegraphics[width=.48\textwidth]{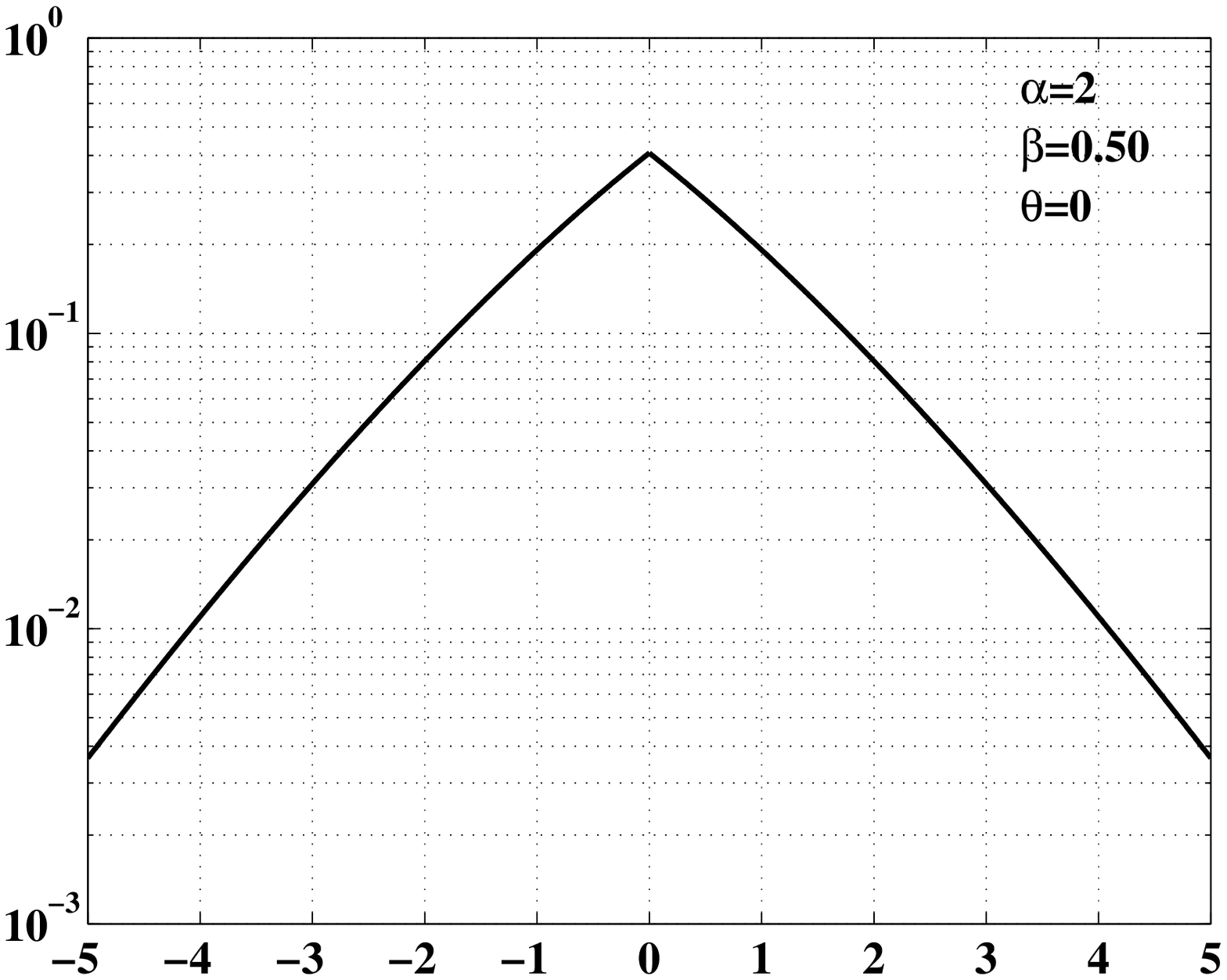}} %%FCAAfig6b.eps
 \vskip 12pt
\caption[]{Probability densities (reduced Green functions)
 %%%for space and time fractional diffusion\\
 for some  values of the triplet $\{\alpha,\theta,\beta\}$}

%% \begin{center}
%% \leavevmode
%% \mpicplace{9 cm}{5 cm}
%% \mpicplace{9 cm}{5 cm}
%% \mpicplace{9 cm}{5 cm}
%% \end{center}

\label{fig1}
\end{figure}

%%%%
% \vfill\eject

%%%%%%%%% THE ELATED STOCHASTIC PROCESSES  %%
As for the  stochastic processes
governed by  the above  probability distributions we can expect the
following.
%%%
For the  case of  non-Gaussian stable densities
we expect  a special class of Markovian processes, called
stable L\'evy motions, which exhibit infinite variance
associated to the possibility of arbitrarily large jumps
({\it L\'evy flights}), whereas for the case   of Wright-type densities
we expect a  class of stochastic non-Markovian processes, which
 exhibit a (finite) variance consistent with slow anomalous diffusion.
Finally, for the case of fractional stable densities,
the related stochastic processes
are expected to possess the  characteristics of the previous two classes.
Indeed,  they are non-Markovian (being $\beta < 1$)
and exhibit infinite variance
associated to the possibility of arbitrarily large jumps
(being $\alpha <2$).
%%%%%%%%%%
A way to realize (understand) all the above stochastic processes is
to show sample paths and histograms of related random walk models.
%% The interested reader can find this for classes of
For random walks discrete both in space and in time  we refer to our
papers
\cite{GorDFMai PhysA99,GorMai NLD02,GorMai PhysA02,GorMai CHEMPHYS02}.

%%%%%%% The END of the Section 3 %%%%

%%%%%%%%%% SECTION 4 %%%%%%%%%%%%%%
\section{From CTRW to fractional diffusion}

Here we show how the space-time fractional diffusion equation (2.2)
can be obtained from the master equation for a
{\it continuous time random walk}
or, equivalently, from the master equation
describing a cumulative {\it renewal process}, through an appropriate limit.
As a matter of fact this limit will be   carried out
in the Fourier-Laplace domain, so the corresponding
convergence is to be intended in a weak form, that
is sufficient %%% (not  restrictive)
for our purposes.
%%%
For the basic principles of
{\it continuous time random walk} (simply referred to as CTRW),
that was formerly introduced in Statistical Mechanics  by Montroll and
Weiss \cite{MontrollWeiss 65}, see \eg
\cite{%% Balescu 97,%%
Hughes 95,%%
%% MontrollScher 73,%%
MontrollShlesingher 84,%%
MontrollWest 79,%%
Weiss 94b},
%% WeissRubin 83
of renewal processes, see \eg
\cite{Cox RENEWAL67,%%
%% Feller RENEWAL41,
Feller 71,%%%
Kotulski 95a,%% Smith RENEW54,Smith RENEW55,
Smith RENEW58}.
%%%%%%

%% The continuous time random walk (CTRW)
The CTRW arises by a sequence
of  independently identically  distributed ($iid$)
 positive  random  waiting times $T_1, T_2, \dots ,$
each having a $pdf$ %% probability density function
 $\psi(t)\,,$   $\, t>0\,, $ and
a sequence of $iid$ random jumps $X_1, X_2, X_3,....$
in $\RR\,,$ each having a $pdf$
$w(x)\,,$ $\, x\in \RR\,.$
%% with $ w(x) = w(-x)\,. $
Setting
$t_0=0\,,$ $\, t_n = T_1+T_2 + \dots T_n$ for $n \in \NN\,,$
$0< t_1<t_2 < \dots\,,$
the wandering particle starts at point $x=0$
in instant $t=0$ and makes a jump
of length $X_n$ in instant $t_n$,
so that its position is
$$x=0 \q \hbox{for} \q
0\le t <T_1= t_1\,,$$   %%%       and
$$ x= S_n =    X_1 + X_2 + \dots X_n\,, \q
\hbox{for} \q t_n \le t < t_{n+1}\,. $$
An essential assumption  is that the waiting time  distribution
and  the jump width distribution  are independent
of each other.
%%%%%%%
It is well known that this stochastic process is {\it Markovian}
if and only if the waiting time $pdf$ is of the form
$\psi(t) = m \, \exp {(-mt)}$ with some
positive constant $m$ ({\it compound Poisson process}),
see \eg \cite{Feller 71}.
%%%
Then, by natural probabilistic arguments we arrive at the
{\it master equation}
 for the spatial $pdf$ %% probability density function ($pdf$)
$p(x,t)$
of the particle being in point $x$ at instant $t\,, $
see  \cite{Gorenflo KONSTANZ01,Mainardi BONN00,Scalas 00},
$$
   p(x,t) =  \delta (x)\, \Psi(t) +
   \int_0^t   \psi(t-t') \, \l[
 \int_{-\infty}^{+\infty}  w(x-x')\, p(x',t')\, dx'\r]\,dt'
 \,, \eqno(4.1) $$
 in which $\delta (x)$ denotes the Dirac generalized function, and,
for abbreviation,
$ \Psi(t) = \int_t^\infty \psi(t') \, dt'\,, $
is the probability that at instant $t$ the particle
is still sitting in its starting position
$x=0\,. $
For this reason  the function $ \Psi(t) $ is usually referred
to as the {\it survival probability}; in the
{\it Markovian} case it reduces to the exponential
function $ \Psi(t) = \exp{(-mt)}\,. $
Actually, $p(x,t)$ as containing a point measure, is a generalized
$pdf$, but for ease of language we omit the qualification
"generalized".
Clearly, (4.1) satisfies the initial condition
$p(x,0^+) = \delta (x)\,. $%%%%

It is customary (and convenient for our purposes) to
consider the master equation (4.1)
in the Fourier-Laplace domain%%%
%%% FOOTNOTE on the HISTORY on CTRW
\footnote{
It was in this domain that originally in 1965 Montroll and Weiss
\cite{MontrollWeiss 65} derived their celebrated equation
for the  CTRW in Statistical Mechanics.
However such equation can be derived
by simply considering a random walk subordinated
to a time renewal process as noted by us in \cite{GorMai LEVY02}
and by Baeumer and  Meerschaert
in  \cite{BaeumerMeerschaert 01}.
%%  see their final remark, p. 498):  a tractable model with
%% time correlation is the continuous time  random walk (CTRW),
%% a simple random walk subordinated to a renewal process.
%% FM: form the more old theory  of renewal processes in the framework
%% of Stochastic Processes
}  %% THE END OF FOOTNOTE ON CTRW
where it reads %%   appears as
$$ \widehat{\widetilde p}(\kappa ,s)
  =  \widetilde {\Psi}(s)   +
   \tilde\psi(s)  \,
 \widehat w(\kappa )\, \widehat{\widetilde p} (\kappa,s)\,,
\eqno(4.2)$$
whence,
$$ \widehat{\widetilde p}(\kappa ,s)
  =  {\widetilde {\Psi}(s)   \over
  1- \widehat w (\kappa )\, \widetilde \psi(s)}
 = {1-\widetilde\psi(s)  \over s}\,
 {1 \over 1- \widehat  w(\kappa )\,\widetilde \psi(s)}\,.
 \eqno(4.3) $$
%% see \eg .......

We will henceforth assume that in our continuous time random walk
the jump width $pdf$ $w(x)$  is an even function
($w(x) = w(-x)$) and has a finite second moment (variance)
or   exhibits the asymptotic  behaviour
$w(x) \sim b\,|x|^{-(\alpha +1)}$
with some $\alpha\,,$ $\,0< \alpha  <2\,,$ for $|x| \to \infty\,, $
and the waiting time $pdf$ $\psi(t)$ has a finite first moment
(mean) or exhibits the asymptotic behaviour
%% obey asymptotic power laws:
$\psi(t) \sim c\, t^{-(\beta +1) } $
with some $\beta\,,$ $\,0< \beta <1\,,$
for $t \to \infty\,,$
where $ b$ and $c$ are  positive constants.
%% In the case $\alpha =2$ we will simply require that
%% $ w(x)$ has a finite variance.
%% In the special (Markovian) case
%% $\beta =1\,,$   which we will
%% briefly discuss at the end, we will take $\psi(t) = \e^{-t}\,.$

%%%%%%%%%%%%%%
Our aim is to derive
from the master equation (4.1),
by  properly rescaling the
waiting times and the jump widths and passing to the diffusion limit,
the {\it space-time fractional diffusion equation}.
By our derivation of (2.2) from (4.1) we de-mystify the often
asked-for meaning of the time fractional derivative
in the fractional diffusion equation.
In plain words, the fractional derivatives in time as well as in space
are caused by asymptotic power laws and
well-scaled passage to the diffusion limit.

Scaling is achieved by making smaller all waiting times by a
positive factor $\tau \,, $  and all jumps
by a positive factor $h\,.$
So we get the jump instants
$$
t_n(\tau ) = \tau T_1 + \tau T_2 +
 \dots + \tau T_n \q \hbox{for} \q n \in \NN\,, \eqno(4.4)$$
and the jump sums,
$$ S_0(h )=0\,, \q
 S_n(h) = h X_1 + h X_2 +
 \dots + h X_n \q \hbox{for} \q n \in \NN\,. \eqno(4.5)$$
The reduced waiting times $\tau T_n$ all have the $pdf$
$\psi_\tau (t) = \psi(t/\tau )/\tau \,, \, t>0\,, $
and analogously the reduced jumps
$hX_n$ all have the $pdf$
$w_h(x) = w(x/h)/h \,, \, x \in \RR\,. $
Readily we see
$$ \widetilde \psi_\tau (s) = \widetilde\psi(s\tau )\,,\q
   \widehat w_h (\kappa) = \widehat w(\kappa h )\,.
\eqno(4.6)$$
Replacing in (4.1)
$\psi(t)$ by $\psi_\tau (t)\,, $
$ \Psi(t)$ by
$ \Psi_\tau (t)= \int_t^\infty \psi_\tau (t')\, dt'\,,$
$w(x)$ by $w_h(x)\,,$
$p(x,t)$ by $p_{h,\tau } (x,t)$
we obtain the rescaled master equation which in the
Fourier-Laplace domain reads as
$$    \widehat{\widetilde  {p_{h,\tau}}}(\kappa ,s) =
   {1 - \widetilde \psi_\tau (s)  \over s}
+   \widetilde \psi_\tau (s) \, \widehat w_h(\kappa )\,
      \widehat{\widetilde  {p_{h,\tau}}}(\kappa ,s) \,, \eqno(4.7)$$
whose solution is
$$    \widehat{\widetilde  {p_{h,\tau}}}(\kappa ,s) =
      {1-\widetilde\psi_\tau (s)  \over s}\,
 {1 \over 1- \widehat  w_h(\kappa )\,\widetilde \psi_\tau (s)}\,.
 \eqno(4.8) $$
%%%%%%%
\vskip 4pt

To proceed further we assume the probability densities  $w(x)$
and $\psi(t)$ of the jumps $X_n$ and the waiting times $T_n$
to meet the asymptotic conditions of the following Lemma 1 and Lemma 2,
respectively,
herewith recalled %%   (without proof)
from \cite{GorMai LEVY02}
where the interested reader can find the proofs.
%% For their proof we refer to the Appendix.

The first Lemma
is a modified specialisation
of Gnedenko's theorem in \cite{GnedenkoKolmogorov 54},
see also \cite{ChechkinGonchar PhysA00}.
It was already used by us, but not formally called a Lemma,
in \cite{GorMai CHEMNITZ01}.
The second Lemma can be obtained
by aid of a corollary in Widder's book \cite{Widder 46}.
%%%%%%%%%%\vfill\eject%%%%%%%%%%%
\\
\vskip 8pt
\noindent{\underbar{\it Lemma 1}}
\\
\vskip 1pt
\noindent
Assume $w(x)\ge 0 \,,$  $w(x) = w(-x)$ for $x \in \RR\,,$
$  \int_{-\infty}^{+\infty}  w(x)\, dx  =1\,,$
and either
$$  \sigma ^2 := \int_{-\infty}^{+\infty}  x^2\, w(x)\, dx <\infty
 \eqno(4.9)$$
(relevant in the case $\alpha =2$) or,
with $b>0$
and some $\alpha \in (0,2)\,, $
$$ w(x) = \l( b +\epsilon (|x|) \r)|x|^{-(\alpha +1)}\,.
  \eqno(4.10)$$
In (4.10) assume
$\epsilon (|x|)$ bounded and
$ O \l(|x|^{-\eta}\r)\,$ with some $\eta >0$ as $    |x|\to   \infty\,.$
%%%%
Then, with a positive scaling parameter $h$ and
a scaling constant
$$ \mu = \cases{
     {\ds {\sigma ^2\over 2}}\,,  \;& if $\q \alpha =2 \,,$ \cr
    {\ds {b\,\pi \over \Gamma(\alpha +1)\,\sin(\alpha \pi/2)}}\,,\;
    & if $\q 0<\alpha <2 \,,$ \cr}
\eqno(4.11)$$
we have, for each fixed $\kappa \in \RR\,,$ the asymptotic relation
%%%%
$$  \widehat w(\kappa h) =
1 -\mu\, (|\kappa|  \, h)^\alpha + o (h^\alpha ) \q
\hbox{for} \q h \to 0\,. \eqno(4.12)$$
%% \underbar{\it Remark}
We note that  (4.12) holds trivially if $\kappa =0$ since
$\widehat w(0)=1\,. $

%%%%%%%%%%%%%%
% \vfill\eject
%%%%%%%%%%%%%\vskip 4pt
\noindent{\underbar{\it Lemma 2}}
\\
\vskip 1pt
\noindent
Assume $\psi(t) \ge 0$ for $t>0\,,$   $\int_0^\infty \psi(t)\,dt =1\,, $
and  either
 $$ \rho := \int_0^\infty t\, \psi(t)\, dt < \infty \eqno(4.13)$$
(relevant in the case $\beta =1$), or,
with $c>0 $ and some $\beta \in (0,1)\,, $
$$ \psi(t) \sim c\, t^{-(\beta +1)}
 \q \hbox{for} \q t \to \infty\,. \eqno(4.14)$$
Then, with a positive scaling parameter $\tau $
and  a scaling constant
$$ \lambda  = \cases{
     {\ds \rho }\,,  \;& if $\q \beta  =1 \,,$ \cr\cr
    {\ds { c\, \Gamma(1-\beta) \over \beta }}\,,\;
    & if $\q 0<\beta  <1 \,,$ \cr}
\eqno(4.15)$$
we have, for each fixed $s>0\,,$  the asymptotic relation
$$ \widetilde \psi(s\tau ) =
     1 - \lambda \, (s\tau )^\beta
 + o(\tau^\beta ) \q \hbox{for} \q \tau  \to 0\,. \eqno(4.16)$$
We note that  (4.16) holds trivially if
$s =0$ since $\widetilde \psi(0) =1$.
\\
\vskip 4pt
\par
%%%%%%%%%%%%%
Eq. (4.8) then becomes asymptotically
  $$    \widehat{\widetilde  {p_{h,\tau}}}(\kappa ,s) \sim
      {\lambda \,\tau^{\beta}\, s^{\beta -1}
\over \lambda \,\tau^{\beta}\, s^{\beta} +
\mu \, h^\alpha \, |\kappa |^\alpha }\,,
\q \hbox{for} \q h,\tau  \to 0 \,.  \eqno(4.17)$$
By imposing the {\it scaling relation}
$$   \lambda \,\tau^{\beta} = \mu \, h^\alpha \,, \eqno(4.18)
 $$
the asymptotics (4.17) yields
$$    \widehat{\widetilde  {p_{h,\tau}}}(\kappa ,s) \to
{ s^{\beta -1} \over s^\beta  + |\kappa |^\alpha} \,.
 \eqno(4.19) $$
%%%%%
Hence, in view of (2.18),
$$\widehat{\widetilde  {p_{h,\tau}}}(\kappa ,s) \to
   \widehat{\widetilde  {\Gzero}}(\kappa ,s) \q
\hbox{for} \q h,\tau  \to 0  \,, \eqno(4.20)$$
under condition (4.18).
%%%%%%
Then, the asymptotic equivalence
in the space-time domain
between the master equation (4.1) after rescaling
and the fractional diffusion equation (2.2) with $\theta =0$
and the initial condition $u(x, 0^+)=\delta (x)$
 is provided
by the continuity theorem for sequences of characteristic functions
after having applied the analogous theorem for sequences of Laplace
transforms, see \eg \cite{Feller 71}.
Therefore  we have {\it convergence in law} or
{\it weak convergence}
for the corresponding probability distributions.
%%%%%%%%%%%% \vfill\eject%%%%%%%%%%%%%

 \section{Simulations}  %%    suggestions}

By aid of the results of Section 4 we can produce
approximate particle paths for space-time fractional diffusion
in the spatially symmetric case $\theta =0$ of (2.2).
To this end, we require, for given $\alpha $ and $\beta $,
a jump width $pdf$ $w(x)$, obeying Lemma 1, and
a waiting time $pdf$ $\psi(t)$, obeying Lemma 2.
Natural choices are the corresponding symmetric stable density
of index $\alpha $, \ie $w(x) = L_\alpha ^0 (x)$ ($0<\alpha \le 2$)
and, following \cite{Mainardi BONN00},
  the corresponding Mittag-Leffler type function
$$
 \psi(t) =  %%%  -   {d \over dt} \Psi(t) =
            -   {d \over dt}  E_\beta (-t^\beta)
 \,, \q 0<\beta \le 1\,,\eqno (5.1)$$
where
$$ E_\beta (z) :=
    \sum_{n=0}^{\infty}\,
   {z^{n}\over\Gamma(\beta\,n+1)}\,, \q \beta >0\,, \q z \in \CC\,,
 \eqno  (5.2)$$
denotes the (entire) transcendental function, known as
the Mittag-Leffler function of order $\beta\,$
\cite{Erdelyi HTF} (Vol. 3, Ch 18, pp. 206-227).
This function, which is a natural generalization of the
exponential  to which it reduces as $\beta =1\,, $
is playing a fundamental role in the applications  of
fractional calculus, see \eg
\cite{GorMai CISM97,Mainardi CISM97}.
%%%%
As has been shown in \cite{Mainardi BONN00},
see also \cite{Hilfer 00a,HilferAnton 95},
this choice  of waiting-time density leads
from the master equation (4.1) to the equation
$$ _tD_*^\beta \,  p(x,t) =
     -  p(x,t) +   \int_{-\infty}^{+\infty} w(x-x')\,
   p(x',t) \, dx'\,, \q p(x,0^+) = \delta(x)\,,
\eqno(5.3) $$
from which, by an appropriately scaled limiting process
(analogous to that of Section 4),
the fractional diffusion equation (2.2) with $u(x,0^+) =  \delta (x)\,,$
can be deduced, see  \cite{Gorenflo  KONSTANZ01}.
Observe that (5.3) in the particular case $\beta =1$
reduces to the well-known Feller-Kolmogorov equation
for a compound Poisson process, in accordance with
$E_1(-t) = \exp (-t)\,. $
%%%%%%%%%%

Still some work must be invested in the inversions
of the cumulative function
$W(x) = \int_{-\infty}^x   w(x')\, dx'$ and
the survival probability
$\Psi(t) = \int_{t}^{\infty}  \psi(t')\, dt'\,,$
which here is
$$ \Psi(t) =
E_\beta (-t^\beta )\,, \q t \ge 0\,, \q 0<\beta \le 1\,.
\eqno(5.4)$$
%%%
In Fig. 2 we exhibit plots of $\Psi(t)$ versus time for some values
of $\beta \in (0,1]$ %% ($\beta = 0.25, 0.50, 0.75, 1$)
from which
we can get insight into the different behaviour for
$0<\beta <1$ (fast decay for short times and slow decay for long times)
and for $\beta =1$ (exponential decay).

%% \section{Figures}
\begin{figure}  %% [b]
\begin{center}

\includegraphics[width=.70\textwidth]{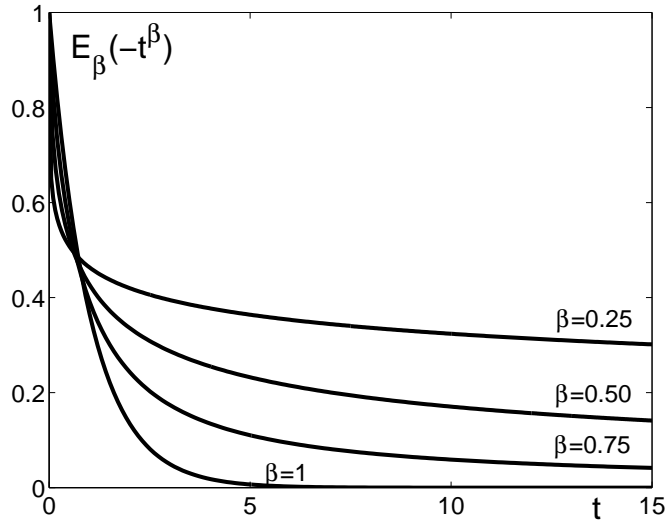}
\end{center}
\caption[]{The survival probability %% Mittag-Leffler function
$\Psi(t) = E_\beta (-t^\beta)$ for
 $\beta = 0.25, 0.50, 0.75, 1$}
\label{fig2}
%% \label{eps1}
\end{figure}

These inversions are required by  the standard Monte-Carlo
procedure of generating the corresponding jump-widths and
waiting-times from $[0,1]$ - uniformly distributed
(pseudo-) random numbers.
%%%%%%%%%%%%%
A. Vivoli in his thesis \cite{Vivoli THESIS02}  has described in detail
how all this can be done and has carried out several case studies
of which we show  (here) three samples for  CTRW's, just to convey a visual
impression on the structure of such processes, see Fig. 3.
In these samples we have $\alpha =2$ so the jump density is
a Gaussian, whereas $\beta =1, 0.75, 0.50\,.$
%%%%%%%%%%
% \vfill\eject
%%%%%%%%%%%%

% \section{Figures}
\begin{figure}  %% [b]
\begin{center}
\includegraphics[width=.70\textwidth]{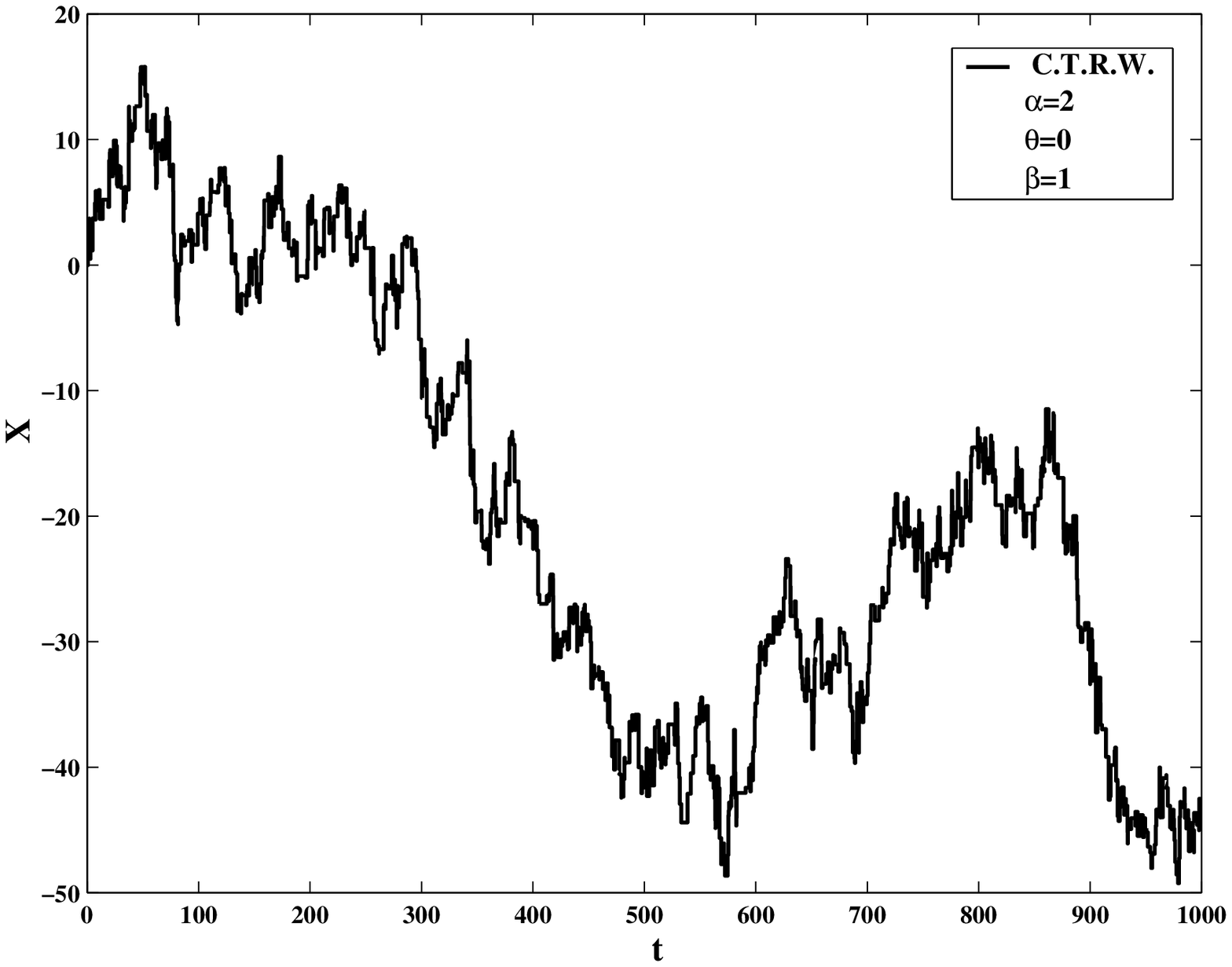} %% AVFIG3_46.eps
\includegraphics[width=.70\textwidth]{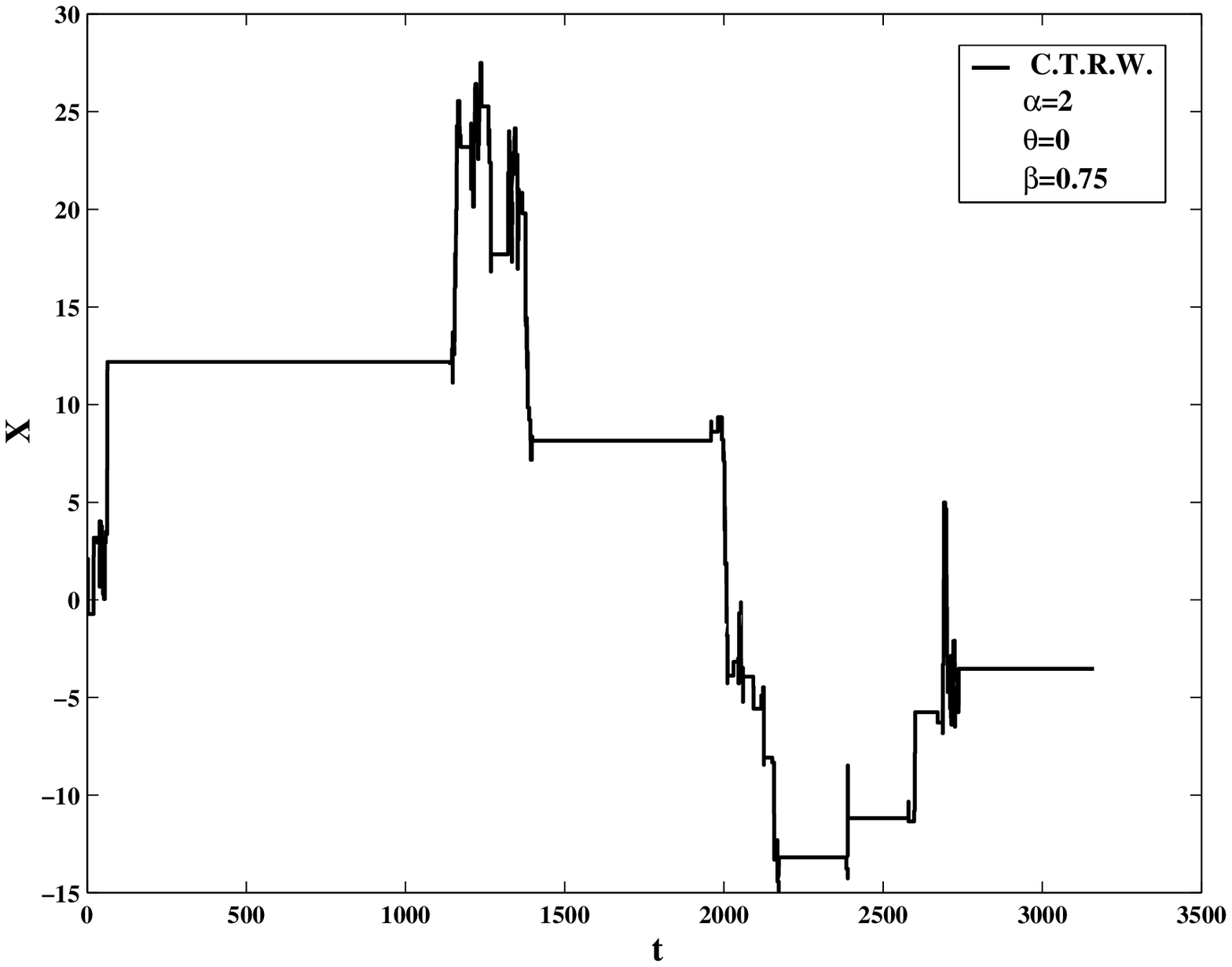} %% AVFIG3_61.eps
\includegraphics[width=.70\textwidth]{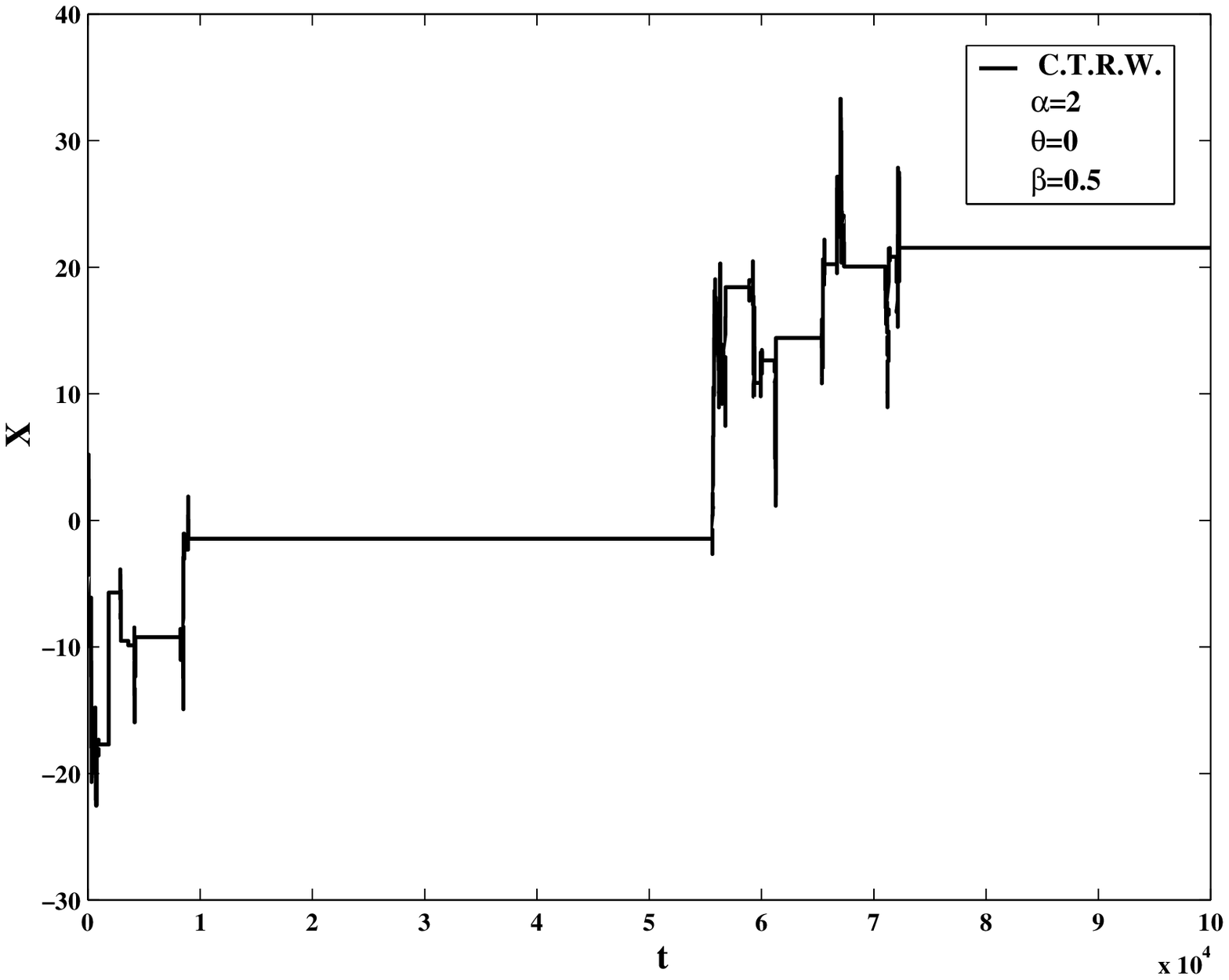} %% AVFIG3_56.eps
\end{center}
\caption[]{Sample paths for  CTRW's with $\alpha =2$,
 $\theta=0$ and $\beta =1, 0.75, 0.50$ \\ (from top to bottom)}
\label{fig3}
%% \label{eps1}
 \end{figure}

%% Please consult the file or printout of \emph{1readme}.$^*$ to find
%% detailed instructions on how to treat figures.

%% If your figures are available as electronic data it is advisable to
%% convert them to eps-format and include them directly into the text with
%% the help of the graphicx package (see the following example in
%% Fig.~\ref{eps1})

%% In order to mark the desired amount of space for a (centered!) figure
%% which has to be pasted into the manuscript manually please provide a
%% vertical line on the lefthand side of the figure. This
%% is reached by using the commands
%% \verb|\mpicplace{width in cm}{height in cm}|.

%% For further instructions
%% e.g. on the structure and layout of the caption see
%% the captions of Figs.~\ref{partfig1} and~\ref{labelfig1}.

Observe in Fig. 3 the striking contrast between the first
graph and the other two. In the case $\beta =1$ we have
$\Psi(t)= \exp(-t)$ which results in long waiting times
occurring rarely    (the mean waiting time being finite!).
So, we get a good approximation of Brownian motion,
(2.2) reducing to (2.1).
%% In the case
For $0<\beta <1$, however,
the Mittag-Leffler function exhibits a power-law decay,
namely
$$ \Psi(t)= E_\beta (-t^\beta )  \, \sim \,
 {\ds {\sin \,(\beta \pi)\over \pi}}
  \,{\ds  {\Gamma(\beta)\over t^\beta}}\,,
  \q  t\to \infty \, .
     \eqno(5.5) $$
As a consequence, we have a distinctly visible preponderance
of long waiting times (the mean waiting time being infinite!).

As our emphasis in this paper is on waiting times (relevant in
CTRW's) we should say that the essential aspect is the asymptotic
behaviour for $t \to \infty$ of the corresponding probability densities,
namely, according to Lemma 2, their decay like $c \, t^{-(\beta +1)}$
($0<\beta <1$) which implies for the survival probability
a decay like $(c/\beta ) \, t^{-\beta }\,. $
This is true, of course, for the Mittag-Leffler
waiting-time distributions used here, see (5.1) and (5.5).
However, in the interest of easy inversion of $\Psi(t)$,
it is advantageous to look
for simpler suitable functions. One such function
(which is more easily invertible) is %% the function
$$ \Psi_*(t) = {1\over  1+ \Gamma(1-\beta) t^\beta } \,,
 \q t\ge 0\,, \q 0<\beta <1\,,
\eqno(5.6)$$
so that
$$ \Psi_*(t) (t) \,\sim
  {\ds {\sin \,(\beta \pi)\over \pi}}
  \,{\ds  {\Gamma(\beta)\over t^\beta}}\,,
  \q  t\to \infty \, .
   \eqno(5.6)  $$
%%%%%%%%
% \vfill\eject
%%%%%%%%%%
\noindent
Happily this function  shares with the function
$\Psi(t) = E_\beta (-t^\beta )$ the desirable property
of complete monotonicity in $t >0\, $
\footnote{    %%% FOOTNOTE ON COMPLETE MONOTONICITY
Complete monotonicity of  a   function
 $f(t)$, $t > 0$, means
$ (-1)^n {\ds{d^n\over dt^n}}\, f  (t) \ge 0\,, $
$\,(n=0,1,2,\dots)$,   a characteristic property of $\exp(-t).$
For the Bernstein theorem  this is
equivalent to the representability of $f(t)$ as (real) Laplace transform
of a given non-negative  (ordinary or generalized)
function.
For more information, see \eg \cite{BergForst 75} (pp. 61-72),
\cite{Feller 71} (pp. 335-338), \cite{Jacob PDOMP01} (pp. 162-164)
and \cite{MillerSamko 01}.
}.
Furthermore we note that the functions $\Psi(t)$ and $\Psi_*(t)$
share the same order of asymptotics for $t \to 0^+$
(albeit with a different coefficient). In fact
we find   as $t \to 0^+\,,$
$$  \Psi(t) = 1- {t^\beta \over \Gamma(1+\beta )} + o(t^\beta )\,,
  \q   \Psi_*(t)= 1-   { \beta \, \pi \over \sin \,(\beta \pi)}\,
{t^\beta \over \Gamma(1+\beta )} +  o(t^\beta )\,.
%%  \q  t \to 0^+ \,.
\eqno(5.7)$$
In a forthcoming paper \cite{GorMai POISSON02}
we will describe in more detail our methods of simulation
and investigate their quality.
%%%
In the interest of long-time (or, because of self-similarity,
near-the-limit) simulations,
it is highly desirable that such fast methods are developed.

 \section{Conclusions}

In this paper we have surveyed the general theory of the one-dimensional
space-time fractional diffusion equation and have presented
representation of its fundamental solutions (the probability densities)
in terms of Mellin-Barnes integrals.
Then, we have outlined how, in the spatially symmetric case,
this equation can be obtained by a limiting process
from a master equation
for a continuous time random walk
via properly scaled compression
of waiting times and jump widths.
For the strictly space and/or time fractional cases
($\{0<\alpha <2\,,\, 0<\beta <1\}$),
 it suffices to assume asymptotic power laws
$ b\, |x| ^ {-(\alpha +1)}$ as $|x| \to \infty$
for the jump width $pdf$  and
$c\, t^{-(\beta +1)}$
as $t \to \infty$  for the waiting time $pdf$.
For the compression factors $h$ in space and $\tau $ in time
we require a scaling relation of the kind
$\lambda \, \tau ^\beta = \mu \, h^\alpha $ where $\lambda ,\mu $ are
given positive constants.
Here we have limited ourselves to show sample paths for some  cases
of the time fractional diffusion processes
(the jump width $pdf$ is Gaussian),
 referring for  more comprehensive
numerical studies to a forthcoming paper.
The theory can be  generalized to more than one
space dimension    and to non-symmetric jump $pdf$'s,
likewise to probability distribution functions instead of densities
for the jump widths and waiting times,
but, in order to avoid too cumbersome notations and calculations,
let us just hint here to such possibilities.

%%%%%%%%%%%%%%%%%%%%%

%%%%%%%%%%%%%%\begin{ack}

%% This part is devoted to acknowledge someone/something ....
%% This work was supported in part by the Italian CNR and INFN,
%% and by the Research Commission of Free University of Berlin.

\vskip 0.4 truecm
\noindent
{\bfs{Acknowledgements}}

\vskip 0.2 truecm
\noindent
We are grateful
to the Erasmus-Socrates project, to the INTAS project  00-0847,
and to the Research Commissions of the Free University of Berlin
%% (Convolution Project)
and of the University of Bologna  %%  (MIUR funds)
for supporting  our work.
%% joint efforts of our research groups in Berlin and Bologna.
%% Our paper is one of the fruits of this collaboration.
F.M. is also grateful to the National Group of Mathematical Physics
(G.N.F.M. - I.N.D.A.M.) and the National Institute of Nuclear Physics
(I.N.F.N. - Sezione  di Bologna) %%  (Theoretical Group).
for partial support.

%% \end{ack}

%% \section{Lists}
%% We have redefined the {\it itemize} environment (labelitemi) so that you
%% will receive bullets instead of dashes to introduce the individual items.
%% We think that this way the list
%% \begin{itemize}
%% \item
%% is clearer
%% \item
%% looks better
%% \item
%% is more noticeable
%% \end{itemize}

%%%%%%%%%
% \vfill\eject
%%%%%%%%%%%%

%INDEX%%%%%%%%%%%%%%%%%%%%%%%%%%%%%%%%%%%%%%%%%%%%%%%%%%%%%%%%%%%%%%%
% Please check with the editor of your book whether he plans to
% include a "mutual" subject index - if so, please code your entries
% in the standard syntax. For your own purposes you may print your
% "personal" index by using the following commands:
%
%\clearpage
%\addcontentsline{toc}{section}{Index}
%\flushbottom
%\printindex
%%%%%%%%%%%%%%%%%%%%%%%%%%%%%%%%%%%%%%%%%%%%%%%%%%%%%%%%%%%%%%%%%%%%%

\end{document}